\documentclass[letter,scriptaddress,twocolumn, prl,showkeys]{revtex4}
\usepackage{amssymb}
\usepackage{amsmath}
\usepackage{makeidx}
\usepackage{amsfonts}
\usepackage{graphicx,epstopdf}
\usepackage[ansinew]{inputenc}
\usepackage[usenames,dvipsnames]{pstricks}
\usepackage{subfigure}
\usepackage{pdfpages}
\usepackage{epsfig}
\usepackage{pst-grad}
\usepackage{pst-plot}
\usepackage[colorlinks,hyperindex]{hyperref}

\hypersetup{
colorlinks,
citecolor=blue,
linkcolor=red,
urlcolor=black,
}

\makeindex

\begin{document}

\title{Creating Oscillons and Oscillating Kinks in Two Scalar Field Theories}
\author{R. A. C. Correa$^{a,b}$ \footnote{%
rafal.couceiro@sissa.it}, A. de Souza Dutra$^{c}$\footnote{%
alvaro.dutra@unesp.br}, T. Frederico$^{b}$\footnote{%
tobias@ita.br}, Boris A. Malomed$^{d}$\footnote{%
malomed@post.tau.ac.il}, O. Oliveira$^{e}$\footnote{%
orlando@uc.pt}, N. Sawado$^{f}$\footnote{%
sawadoph@rs.tus.ac.jp}}
\affiliation{Scuola Internazionale Superiore di Studi Avanzati (SISSA), via Bonomea, 265,
I-34136 Trieste, Italy \\
ITA-Instituto Tecnol\'ogico de Aeron\'autica, 12228-900, S\~ao Jos\'e dos
Campos, SP, Brazil \\
S\~ao Paulo State University (UNESP), Campus de Guaratingueta, 12516-410,
Guaratinguet\'a, SP, Brazil \\
Department of Physical Electronics, School of Electrical Engineering,
Faculty of Engineering, and Center for Light-Matter Interaction, Tel Aviv
University, Tel Aviv 69978, Israel \\
CFisUC, Department of Physics, University of Coimbra, P-3004 516 Coimbra,
Portugal \\
Department of Physics, Tokyo University of Science, Noda, Chiba 278-8510,
Japan}

\begin{abstract}
Oscillons are time-dependent, localized in space, extremely long-lived
states in nonlinear scalar-field models, while kinks are topological
solitons in one spatial dimension. In the present work, we show new classes
of oscillons and oscillating kinks in a system of two nonlinearly coupled
scalar fields in $1 + 1$ spatiotemporal dimensions. The solutions contain a
control parameter, the variation of which produces oscillons and kinks with
a flat-top shape. The model finds applications to condensed matter,
cosmology, and high-energy physics.
\end{abstract}

\maketitle

\textbf{In the context of the field theory it is quite common the appearance
of oscillons, which are time-dependent, localized in space, and extremely
long living scalar field solutions which arise from nonlinear interactions.
On the other hand, kinks are topological solitons in one spatial dimension.
Here, we report the existence of new classes of oscillons and oscillating
kinks in a system of two nonlinearly coupled scalar fields in 1 + 1
space-time dimensions. Such model is of interest in applications that
include condensate matter, cosmology and high energy physics. We show the
existence of diferent kinds of oscillons and oscillating kinks, which are
characterized by the presence of an controllable parameter in the solutions.
As we are changing the value that parameter the oscillons becomes flat in
its top, thereby occupying a significant amount of the spacial distribution.
In the case of the oscillating kinks, we obtain that the behavior of the
solitons changes drastically, by developing a kind of plateau in its center.}

\section{Introduction}

Nonlinear field theories are commonly used in Physics to model phenomena at
all energy scales \cite{ref1}, in condensed matter \cite{ref2}, in quantum
solitons \cite{ref3}, in cosmology \cite{ref4}, in particle physics \cite%
{Bento:2018fmy,Iguro:2018oou,Chakrabarty:2019bbj}, and in other areas of
physics \cite{ref5,ref6,ref7}. One of fundamental reasons for the increasing
interest in nonlinear phenomena is the possibility to have multiple
degenerate minimum-energy configurations of the system. A good testbed to
study the dynamics associated with the presence of degenerated minima are
scalar field models whose potential energy has two or more degenerate
minima. The present work addresses this topic, focusing in a particular
model.

In nonlinear classical-field theories, solitons \cite{ref10,ref8} are
localized field configurations with finite energy density, that keep their
shape unaltered after collision with other solitons. These type of
configurations have been studied in a wide class of contexts, such as
nonlinear lattices \cite{ref11}, Skyrmion models \cite{ref12,ref13},
theories breaking Lorentz invariance \cite{ref14,ref15}, and the
Maxwell-Chern-Simons gauge model \cite{ref16}.

The scalar nonlinear field theories allow for various types of
configurations and, among then, are spatially localized time-periodic states
named oscillons. Oscillons were discovered by Bogolyubsky and Makhankov \cite%
{ref17} and rediscovered \textit{a posteriori} by Gleiser \cite{ref18}.
Typically, they have bell-shaped profiles that oscillates sinusoidally in
time \cite{ref18.1,ref18.2}. Further, oscillons are states in non-integrable
models that continuously loose energy via radiation losses \cite{ref20} but,
nevertheless, can be represented by extremely long-lived modes and preserve
the self-trapping in the respective models \cite{ref21}. These properties
are not shared with other classes of solutions of nonlinear scalar field
theories that the breathers of the one-dimensional sine-Gordon equations 
\cite{ref19} are an example..

A vast body of results have been published for oscillons. In particular,
they have been investigated in the presence of external potentials \cite%
{ref22,Copeland:2014qra}, in string-inspired cosmological models \cite{ref23}%
, in the context of supersymmetry \cite{ref24}, in perturbed sign-Gordon
models \cite{ref25}, and in higher-derivative field theories \cite{ref26}.
Oscillons have also been studied in connection with gravitational waves \cite%
{Zhou:2013tsa,ref27,ref28}, Abelian-Higgs models \cite{ref30}, in the
context of the standard $SU(2) \times U(1)$ theory for electroweak
interactions \cite{ref31}, and in theories probing violation of the Lorentz
and \textit{CPT} symmetries \cite{ref32,ref33}.

An interesting scenario where oscillons can play a fundamental role is after
cosmic inflation \cite{Lozanov:2014zfa, Lozanov:2016pac, Lozanov:2016hid,
Lozanov:2018kpk} as they act as a source of gravitational waves \cite%
{Lozanov:2019ylm} and lead to a reduction in the uncertainties in the
predictions for inflationary observables \cite{Lozanov:2017hjm}.

Oscillons configurations were not studied in detail in models with two or
more interacting scalar fields. In this context, it was pointed out in Ref. 
\cite{ref34} that a hybrid inflation model with two real scalar fields,
interacting quadratically, can produce oscillon configurations whose
lifetimes are much larger than those associated with single-field oscillons.
As shown in that work, such configurations persist for at least four
cosmological expansion times, accounting for up to $20\%$ of the total
energy density of the early Universe.

The oscillons solutions in theories with two scalar fields were also
addressed in Ref. \cite{ref35} in the framework of the dynamics of the
Abelian Higgs model. By using a multiscale expansion method \cite{ref36}, a
class of oscillons and oscillating kinks was obtained for both Higgs and
gauge fields. Moreover, it was argued that similar dynamics occurs in
superconductors and in the superfluid phase of atomic Bose-Einstein
condensates trapped in optical lattices. Furthermore, in the framework of a
similar scenario it was shown \cite{ref37} that it is possible to classify
oscillons solutions according to their mass ratio and, accordingly, they may
correspond to type-I or type-II superconductors. Various fluxon states
produced by a system of sine-Gordon equations modelling a triangular
configuration of three coupled long Josephson junctions were reported in
Ref. \cite{Stan}. It is also relevant to mention that it was recently shown 
\cite{ref49,ref50,ref51} that cosmological backgrounds based on models with
more than one scalar field accurately comply with constraints inferred from
data collected by the Planck satellite.

The aim of the present work is to show that oscillons naturally emerge in a
system of two nonlinearly coupled scalar fields in ($1+1$)-dimensional
spatiotemporal continuum. We address a class of sixth-degree polynomial
potentials, which allows for rich structure of vacua \cite{ref38,ref39}.
This model also plays an important role in the description of topological
twistons in polyethylene \cite{ref40, ref41, ref42} and has also been used
to investigate topological defects in molecular chains with zig-zag \cite%
{ref43} and helix \cite{ref44} structures. Unlike the oscillons and
oscillating kinks considered in Refs. \cite{ref35, ref37}, where solutions
were produced by an effective nonlinear Schr\"odinger equation, herein we
use the expansion procedure developed in Ref. \cite{ref18.1} to construct
two new classes of oscillons and two classes of oscillating kinks. In
particular, it will be shown that oscillating-kink solutions include both
single- and double-kink modes, and oscillons exhibit both bell-shaped and
flat-top profiles.

This paper is organized as follows. In Section 2 we introduce the models
with two interacting scalar fields. In Section 3 the method used for
constructing oscillons solutions is presented. Then, new classes of
oscillons and oscillating kinks are reported in Section 4, and their
applications are discussed in Section 5. The work is concluded in Section 6.

\label{sec:int}


\section{The two-scalar-fields model}

Scalar field models with more than one scalar field naturally emerge in the
study of many physical systems. Systems with two interacting scalar fields,
that we address here, appear as models of networks of BPS
(Bogomol'nyi-Prasad-Sommerfeld) and non-BPS defects \cite{ref45}, are used
to describe fermion localization on degenerate and critical Bloch branes 
\cite{ref45.1,ref46}, on the hierarchy problems \cite{ref47}, and in
field-theory kinks \cite{ref48}.

Here, we aim to construct oscillons and related states in a two-component
system that is close to one considered in Ref. \cite{ref38}. The ($1+1$%
)-dimensional model includes a sixth-degree polynomial potential, 
\begin{eqnarray}
&&\left. V(\phi ,\chi )=\kappa _{1}\phi ^{2}+\kappa _{2}\chi ^{2}+\kappa
_{3}\phi ^{4}+\kappa _{4}\chi ^{4}+\kappa _{5}\phi ^{6}\right.  \notag \\
&&\left. +\kappa _{6}\chi ^{6}+\kappa _{7}\phi ^{2}\chi ^{2}+\kappa _{8}\phi
^{2}\chi ^{4}+\kappa _{9}\phi ^{4}\chi ^{2},\right.  \label{eq0.0}
\end{eqnarray}%
where $\phi =\phi (x,t)$ and $\chi =\chi (x,t)$ are real scalar fields, and $%
\kappa _{i}$ are real coupling constants. Note that with the choice of 
\begin{eqnarray}
\kappa _{1} &=&\kappa _{2}=\frac{\nu ^{2}}{2},\text{ }\kappa _{3}=\mu \nu ,%
\text{ }\kappa _{4}=\lambda \nu ,\text{ }  \label{param} \\
\kappa _{5} &=&\frac{\mu ^{2}}{2},\text{ }\kappa _{6}=\frac{\lambda ^{2}}{2}%
,\kappa _{7}=3\nu (\lambda +\mu ), \\
\kappa _{8} &=&\frac{3\lambda }{2}\left( 3\lambda +2\mu \right) ,\text{ }%
\kappa _{9}=\frac{3\mu }{2}\left( 3\mu +2\lambda \right) ,  \notag
\end{eqnarray}%
$V(\phi ,\chi )$ reproduces the potential energy of the model introduced in 
\cite{ref38}, 
\begin{equation}
V(\phi ,\chi )=\frac{1}{2}\chi ^{2}(\lambda \chi ^{2}+3\mu \phi ^{2}+\nu
)^{2}+\frac{1}{2}\phi ^{2}(3\lambda \chi ^{2}+\mu \phi ^{2}+\nu )^{2}
\label{1}
\end{equation}%
where $\lambda $, $\mu $ and $\nu $ are real positive coupling constants.
This potential has a rich vacuum structure. For instance, for $\nu \lambda
<0 $ and $\nu \mu <0$ it has the following nine minima:

\begin{eqnarray}
\mathcal{M}_{1} &=&\Big(0,0\Big),\text{ }  \notag \\
\mathcal{M}_{2} &=&\Big(-\sqrt{-\frac{\nu }{\mu }},0\Big),  \notag \\
\mathcal{M}_{3} &=&\Big(\sqrt{-\frac{\nu }{\mu }},0\Big),\text{ }  \notag \\
\mathcal{M}_{4} &=&\Big(0,-\sqrt{-\frac{\nu }{\lambda }}\Big),  \notag \\
\mathcal{M}_{5} &=&\Big(0,\sqrt{-\frac{\nu }{\lambda }}\Big),\text{ }
\label{2} \\
\mathcal{M}_{6} &=&\Big(-\sqrt{-\frac{\nu }{4\mu }},-\sqrt{-\frac{\nu }{%
4\lambda }}\Big),  \notag \\
\mathcal{M}_{7} &=&\Big(-\sqrt{-\frac{\nu }{4\mu }},\sqrt{-\frac{\nu }{%
4\lambda }}\Big),\text{ }  \notag \\
\mathcal{M}_{8} &=&\Big(\sqrt{-\frac{\nu }{4\mu }},-\sqrt{-\frac{\nu }{%
4\lambda }}\Big),  \notag \\
\mathcal{M}_{9} &=&\Big(\sqrt{-\frac{\nu }{4\mu }},\sqrt{-\frac{\nu }{%
4\lambda }}\Big).  \notag
\end{eqnarray}%
%
%
%
%
%
%
%
%
where we used the notation $\mathcal{M}_{n}=(\phi ,\chi )$ for two
components of the $n$-th vacuum state.

\begin{figure}[h]
\begin{center}
\includegraphics[width=8.2cm]{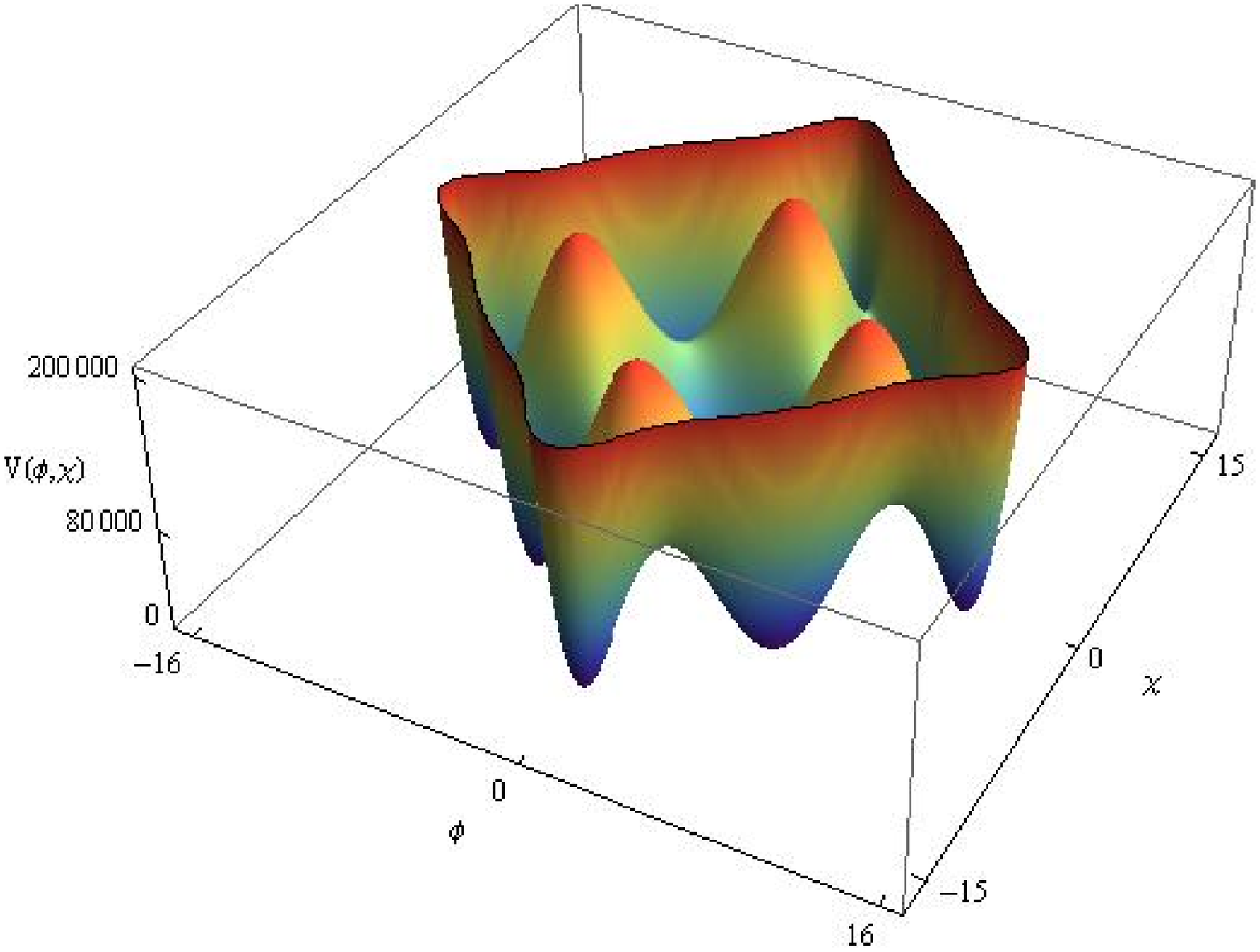} %
\includegraphics[width=8.2cm]{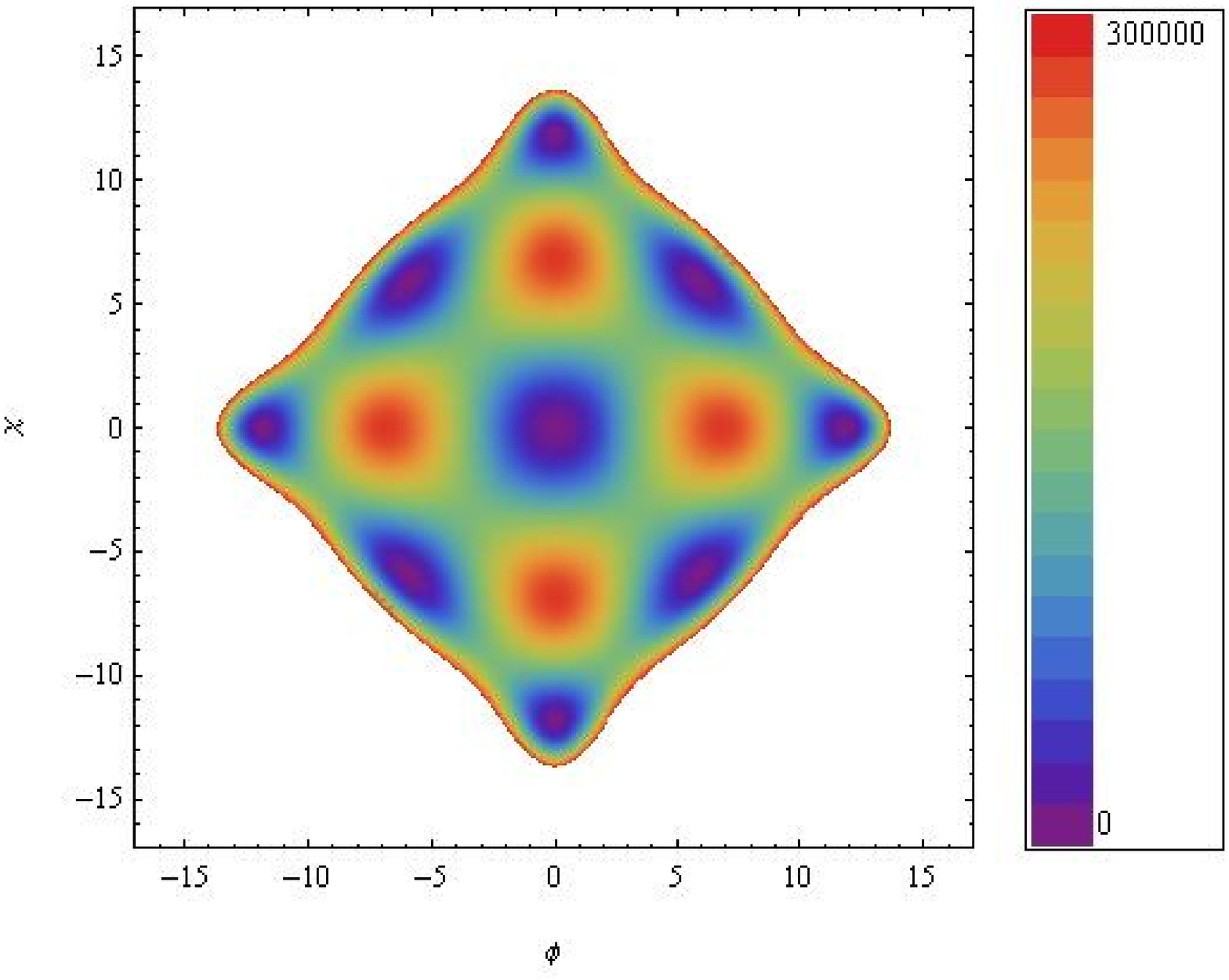}
\end{center}
\caption{The profile of potential $V\left( \protect\phi ,\protect\chi %
\right) $ (top) and vacuum states (botton) produced by Eq. (\protect\ref{1})
with constants $\protect\lambda =\protect\mu =1$ and $\protect\nu =-140.$}
\label{fig1:vacuum states}
\end{figure}

The vacuum structure associated with potential (\ref{1}) is sketched in Fig.~%
\ref{fig1:vacuum states}, that also displays the potential's profile. One
can predict different classical configurations by connecting different vacua.

Previous analysis of this model addressed only static field configurations,
revealing several kink-like and lump-like solutions \cite{ref38}. Our
objective is to investigate solutions of the field equations that are time-
and space-dependent.

\section{Solutions of the coupled field equations}

The two-component system is defined by the Lagrangian density, 
\begin{equation}
\mathcal{L}=\frac{1}{2}(\partial _{\alpha }\phi )^{2}+\frac{1}{2}(\partial
_{\alpha }\chi )^{2}-V(\phi ,\chi ),  \label{3}
\end{equation}%
with $V(\phi ,\chi )$ given by Eq. (\ref{1}). We use a system of units such
that $c=\hbar =1$, the metric is $\eta _{\alpha \beta }=$ diag$(1,-1)$, and
the coordinates are $x^{\alpha }=(t,x)$. The classical equations of motion
generated by $\mathcal{L}$\ are 
\begin{eqnarray}
\frac{\partial ^{2}\phi (x,t)}{\partial t^{2}}-\frac{\partial ^{2}\phi (x,t)%
}{\partial x^{2}}+\frac{\partial V(\phi ,\chi )}{\partial \phi } &=&0,
\label{4} \\
&&  \notag \\
\frac{\partial ^{2}\chi (x,t)}{\partial t^{2}}-\frac{\partial ^{2}\chi (x,t)%
}{\partial x^{2}}+\frac{\partial V(\phi ,\chi )}{\partial \chi } &=&0.
\label{5}
\end{eqnarray}%
\ 

A small perturbation will be added to potential (\ref{eq0.0}) as a means for
creating time-dependent configurations. A simple way to introduce the
perturbation is to replace 
\begin{equation}
\kappa _{1,2}\rightarrow \kappa _{1,2}-\delta ^{2}\epsilon ^{2}
\label{kappa}
\end{equation}%
in Eq. (\ref{eq0.0}), where $\epsilon <<1$ is a small parameter, while $%
\delta $ is a generic coefficient (it is required in this definition as the
same small parameter $\epsilon $ is introduced below in another context, see
Eqs. (\ref{e9}) and (\ref{eps}) ).

Our main objective is to find very long-lived spatially localized
time-dependent configurations, using an appropriate procedure devised in
Ref. \cite{ref18.1}. As a first step, $x$ and $t$ are rescaled as 
\begin{equation}
\tau =\omega ~t,\qquad y=\epsilon ~x,  \label{e9}
\end{equation}%
where $\epsilon $ is the same small parameter as in Eq. (\ref{kappa}), and%
\begin{equation}
\omega \equiv \sqrt{1-\epsilon ^{2}}.  \label{eps}
\end{equation}
Such transformations naturally appear in the context of the multiscale
expansion method \cite{ref36}, which defines several temporal and spatial
variables, that are scaled differently and treated as independent ones.

Under the scale transformation (\ref{e9}), Eqs. (\ref{4}) and Eq. (\ref{5}),
with the potential energy given in Eq. (\ref{eq0.0}), become 
\begin{eqnarray}
&&\left. (1-\epsilon ^{2})\frac{\partial ^{2}\phi }{\partial \tau ^{2}}%
-\epsilon ^{2}\frac{\partial ^{2}\phi }{\partial y^{2}}=-\left( 2(\kappa
_{1}-\epsilon ^{2}\delta ^{2})\phi \right. \right.  \label{8} \\
&&\left. \left. +4\kappa _{3}\phi ^{3}+6\kappa _{5}\phi ^{5}+2\kappa
_{7}\phi \chi ^{2}+2k_{8}\phi \chi ^{4}+4\kappa _{9}\phi ^{3}\chi
^{2}\right) ,\right.  \notag \\
&&  \notag \\
&&\left. (1-\epsilon ^{2})\frac{\partial ^{2}\chi }{\partial \tau ^{2}}%
-\epsilon ^{2}\frac{\partial ^{2}\chi }{\partial y^{2}}=-\left( 2(\kappa
_{2}-\epsilon ^{2}\delta ^{2})\chi \right. \right.  \label{9} \\
&&\left. \left. +4\kappa _{4}\chi ^{3}+6\kappa _{6}\phi ^{5}+2\kappa
_{7}\phi ^{2}\chi +2\kappa _{8}\phi ^{2}\chi ^{3}+4\kappa _{9}\phi ^{4}\chi
\right) .\right.  \notag
\end{eqnarray}

The standard approach used to obtain oscillons configurations is based on
the expansion of the scalar fields as powers of $\epsilon $. Given that Eqs.
(\ref{8}) and (\ref{9}) are odd in the fields, the expansions defining $\phi 
$ and $\chi $ include only odd powers of $\epsilon $: 
\begin{eqnarray}
\phi (y,\tau ) &=&\sum_{n=1}^{\infty }\epsilon ^{2n-1}\phi _{2n-1}(y,\tau ),
\label{10} \\
&&  \notag \\
\chi (y,\tau ) &=&\sum_{n=1}^{\infty }\epsilon ^{2n-1}\chi _{2n-1}(y,\tau ).
\label{11}
\end{eqnarray}%
Then, Eqs. (\ref{8}) and (\ref{9}) yield 
\begin{eqnarray}
&&\left. \epsilon \left( \frac{\partial ^{2}\phi _{1}}{\partial \tau ^{2}}%
+2\kappa _{1}\phi _{1}\right) +\epsilon ^{3}\left( \frac{\partial ^{2}\phi
_{3}}{\partial \tau ^{2}}+2\kappa _{1}^{2}\phi _{3}-\frac{\partial ^{2}\phi
_{1}}{\partial \tau ^{2}}\right. \right.  \notag \\
&&\left. \left. -\frac{\partial ^{2}\phi _{1}}{\partial y^{2}}-2\delta
^{2}\phi _{1}+4\kappa _{3}\phi _{1}^{3}+2\kappa _{7}\phi _{1}\chi
_{1}^{2}\right) +\mathcal{O(\epsilon }^{5})=0,\right.  \notag \\
&&  \label{12} \\
&&\epsilon \left( \frac{\partial ^{2}\chi _{1}}{\partial \tau ^{2}}+2\kappa
_{2}\chi _{1}\right) +\epsilon ^{3}\left( \frac{\partial ^{2}\chi _{3}}{%
\partial \tau ^{2}}+2\kappa _{2}\chi _{3}-\frac{\partial ^{2}\chi _{1}}{%
\partial \tau ^{2}}\right.  \notag \\
&&\left. \left. -\frac{\partial ^{2}\chi _{1}}{\partial y^{2}}-2\delta
^{2}\chi _{1}+4\kappa _{4}\chi _{1}^{3}+2\kappa _{7}\phi _{1}^{2}\chi
_{1}\right) +\mathcal{O(\epsilon }^{5})=0.\right.  \notag \\
&&  \label{13}
\end{eqnarray}

In the two lowest orders, Eqs. (\ref{12}) and (\ref{13}) lead to the
following set of coupled nonlinear differential equations: 
\begin{eqnarray}
&&\left. \frac{\partial ^{2}\phi _{1}}{\partial \tau ^{2}}+2\kappa _{1}\phi
_{1}=0,\right.  \label{14} \\
&&  \notag \\
&&\left. \frac{\partial ^{2}\chi _{1}}{\partial \tau ^{2}}+2\kappa _{2}\chi
_{1}=0,\right.  \label{15} \\
&&  \notag \\
&&\left. \frac{\partial ^{2}\phi _{3}}{\partial \tau ^{2}}+2\kappa _{1}\phi
_{3}=\frac{\partial ^{2}\phi _{1}}{\partial \tau ^{2}}+\frac{\partial
^{2}\phi _{1}}{\partial y^{2}}\right.  \label{16} \\
&&\left. +2\delta ^{2}\phi _{1}-4\kappa _{3}\phi _{1}^{3}-2\kappa _{7}\phi
_{1}\chi _{1}^{2},\right.  \notag \\
&&  \notag \\
&&\left. \frac{\partial ^{2}\chi _{3}}{\partial \tau ^{2}}+2\kappa _{2}\chi
_{3}=\frac{\partial ^{2}\chi _{1}}{\partial \tau ^{2}}+\frac{\partial
^{2}\chi _{1}}{\partial y^{2}}\right.  \label{17} \\
&&\left. +2\delta ^{2}\chi _{1}-4\kappa _{4}\chi _{1}^{3}-2\kappa _{7}\phi
_{1}^{2}\chi _{1}.\right.  \notag
\end{eqnarray}%
Equations (\ref{14}) and (\ref{15}) are of the harmonic-oscillator type,
hence their solution can be written as 
\begin{eqnarray}
\phi _{1}(y,\tau ) &=&\varphi (y)~\cos (\sqrt{2\kappa _{1}}\tau ),
\label{18} \\
&&  \notag \\
\chi _{1}(y,\tau ) &=&\sigma (y)~\sin (\sqrt{2\kappa _{2}}\tau ).  \label{19}
\end{eqnarray}%
Functions $\varphi (y)$ and $\sigma (y)$, i.e. the spatial profile of the
configurations appearing in Eqs. (\ref{18}) and (\ref{19}), can be obtained
inserting $\phi _{1}(y,\tau )$ and $\chi _{1}(y,\tau )$ in Eqs. (\ref{16})
and (\ref{17}). After straightforward manipulations, one gets 
\begin{eqnarray}
&&\left. \frac{\partial ^{2}\phi _{3}}{\partial \tau ^{2}}+2k_{1}\phi
_{3}=\right.  \notag \\
&&\left. -\left( 2(\kappa _{1}-\delta ^{2}){\varphi +3\kappa }_{3}{\varphi
^{3}+\kappa }_{7}{\varphi \sigma ^{2}-}\frac{d^{2}{\varphi }}{dy^{2}}\right)
\cos (\sqrt{2\kappa _{1}}\tau )\right.  \notag \\
&&\left. -\varphi (\kappa _{3}\varphi ^{2}-\frac{\kappa _{7}}{2}\sigma
^{2})\cos (3\sqrt{2\kappa _{1}}\tau ),\right.  \label{20} \\
&&  \notag \\
&&\left. \frac{\partial ^{2}\chi _{3}}{\partial \tau ^{2}}+2k_{2}\chi
_{3}=\right.  \notag \\
&&\left. -\left( 2(\kappa _{2}-\delta ^{2}){\sigma +3\kappa }_{4}{\sigma
^{3}+\kappa }_{7}{\varphi ^{2}\sigma -}\frac{d^{2}{\sigma }}{dy^{2}}\right)
\sin (\sqrt{2\kappa _{2}}\tau )\right.  \notag \\
&&+\sigma (\kappa _{4}\sigma ^{2}-\frac{\kappa _{7}}{2}\varphi ^{2})\sin (3%
\sqrt{2\kappa _{2}}\tau ).  \label{21}
\end{eqnarray}

For field configurations that are periodic in time, the contribution
proportional to $\cos (\sqrt{2\kappa _{1}}\tau )$ and $\sin (\sqrt{2\kappa
_{2}}\tau )$ in the right-hand side of Eqs. (\ref{20}) and (\ref{21}) must
vanish, otherwise solutions of the equations produced by these \emph{resonant%
} driving terms will generate a term linear in $\tau $, hence the solutions
for $\phi _{3}$ and $\chi _{3}$ will be not be time-periodic ones. This
condition translates into the equations 
\begin{eqnarray}
\frac{d^{2}{\varphi }}{dy^{2}} &=&2(\kappa _{1}-\delta ^{2}){\varphi
+3\kappa }_{3}{\varphi ^{3}+\kappa }_{7}{\varphi \sigma ^{2},}  \label{22} \\
&&  \notag \\
\frac{d^{2}{\sigma }}{dy^{2}} &=&2(\kappa _{2}-\delta ^{2}){\sigma +3\kappa }%
_{4}{\sigma ^{3}+\kappa }_{7}{\varphi ^{2}\sigma ,}  \label{23}
\end{eqnarray}%
that determine $\varphi (y)$ and $\sigma (y)$. In general, nonlinear
equations (\ref{22}) and (\ref{23}) are quite difficult to solve. However,
as shown in the next section, they can be mapped into first-order linear
differential equations whose general solution can be constructed by means of
standard methods.

\section{Oscillons and oscillating-kink solutions}

Now, our goal is to reduce the second-order differential equations (\ref{22}%
) and (\ref{23}) to first-order ones. To this end, we write these equations
as 
\begin{equation}
\frac{d^{2}{\varphi }}{dy^{2}}=U_{\varphi }(\varphi ,\sigma ),\qquad \qquad 
\frac{d^{2}{\sigma }}{dy^{2}}=U_{\sigma }(\varphi ,\sigma ),  \label{24}
\end{equation}%
where 
\begin{eqnarray}
U_{\varphi }(\varphi ,\sigma ) &\equiv &2(\kappa _{1}-\delta ^{2}){\varphi
+3\kappa }_{3}{\varphi ^{3}+\kappa }_{7}{\varphi \sigma ^{2},}  \label{25} \\
&&  \notag \\
U_{\sigma }(\varphi ,\sigma ) &\equiv &2(\kappa _{2}-\delta ^{2}){\sigma
+3\kappa }_{4}{\sigma ^{3}+\kappa }_{7}{\varphi ^{2}\sigma .}  \label{26}
\end{eqnarray}%
%
%
%
%
%
%
%
%
%
%
%
To decouple the pair of Eqs. (\ref{24}), we multiply the first and second
equation by $d\sigma /dy$ and $d\varphi /dy$, respectively, and take their
sum, to arrive at the result 
\begin{equation}
\frac{1}{2}\left[ \left( \frac{d{\varphi }}{dy}\right) ^{2}+\left( \frac{d{%
\sigma }}{dy}\right) ^{2}\right] +U(\varphi ,\sigma )=\alpha _{0}.
\label{eq4}
\end{equation}%
Here $\alpha _{0}$ is an arbitrary constant, that should be set equal to
zero to get solutions that connect different vacua of the system, and $%
U(\varphi ,\sigma )$ is an effective potential that can be written, using a
supersymmetric representation for function $U(\varphi ,\sigma )$ in terms of
a \textit{superpotential}, $\mathcal{W}(\varphi ,\sigma )$, as 
\begin{equation}
U(\varphi ,\sigma )=\frac{1}{2}[\mathcal{W}_{\varphi }^{2}\mathcal{(}\varphi
,\sigma )+\mathcal{W}_{\sigma }^{2}\mathcal{(}\varphi ,\sigma )],  \label{27}
\end{equation}%
where subscripts stand for derivatives with respect to $\varphi $ and $%
\sigma $, and the appropriate superpotential is 
\begin{equation}
\mathcal{W(}\varphi ,\sigma )=-a_{1}\varphi +\frac{b_{1}}{3}\varphi
^{3}+c_{1}\varphi \sigma ^{2},  \label{eq7}
\end{equation}%
with the following definitions 
\begin{eqnarray}
a_{1} &\equiv &\sqrt{\frac{2(\delta ^{2}-\kappa _{2})(3\delta ^{2}-\kappa
_{1}-\kappa _{2})}{\kappa _{7}}},\text{ }  \label{eq7.11} \\
b_{1} &\equiv &(\delta ^{2}-\kappa _{1})\sqrt{\frac{\kappa _{7}}{2(\delta
^{2}-\kappa _{2})(3\delta ^{2}-\kappa _{1}-\kappa _{2})}},  \label{eq714} \\
c_{1} &\equiv &(\delta ^{2}-\kappa _{2})\sqrt{\frac{\kappa _{7}}{2(\delta
^{2}-\kappa _{2})(3\delta ^{2}-\kappa _{1}-\kappa _{2})}}.  \label{rq715}
\end{eqnarray}%
%
%
%
%
%
%
%

It follows from Eqs. (\ref{eq7.11})-(\ref{rq715}) that, to produce real
solutions, the constants should be subject to the following restrictions: 
\begin{equation}
\delta ^{2}<\kappa _{1},\qquad \delta ^{2}>(\kappa _{1}+\kappa
_{2})/3,\qquad \kappa _{7}<0,\qquad \kappa _{3}=\kappa _{4}.  \label{eq716}
\end{equation}

The advantage of this transformation is that it is possible to reduce the
second-order differential equations (\ref{24}) to first-order ones, 
\begin{equation}
\frac{d\varphi }{dy}=\mathcal{W}_{\varphi },\qquad \qquad \frac{d\sigma }{dy}%
=\mathcal{W}_{\sigma }.  \label{28}
\end{equation}%
Following Ref. \cite{ref52}, one obtains from Eq. (\ref{28}) an equation for 
$\varphi $ considered as a function of $\sigma $: 
\begin{equation}
\frac{d\varphi }{d\sigma }=\frac{\mathcal{W}_{\varphi }}{\mathcal{W}_{\sigma
}}=\frac{b_{1}(\varphi ^{2}-1)+c_{1}\sigma ^{2}}{2c_{1}\varphi \sigma }.
\label{29}
\end{equation}%
In terms of variable $\rho \equiv \varphi ^{2}-1$, Eq. (\ref{29}) is cast in
the form of a linear ODE,%
\begin{equation}
\frac{d\rho }{d\sigma }-\frac{b_{1}\rho }{c_{1}\sigma }=\sigma ,  \label{30}
\end{equation}%
whose general solution is 
\begin{eqnarray}
\rho (\sigma ) &=&c_{0}\sigma ^{b_{1}/c_{1}}-\frac{c_{1}}{b_{1}-2c_{1}}%
\sigma ^{2},\qquad (b_{1}\neq 2c_{1})  \label{37} \\
\rho (\sigma ) &=&\sigma ^{2}[\ln (\sigma )+d_{1}],\qquad (b_{1}=2c_{1}),
\label{38}
\end{eqnarray}%
where $c_{0}$ and $d_{1}$ are arbitrary integration constants. Next, these
solutions feed the first-order ODE for $\sigma $ in Eq. (\ref{28}), that
become 
\begin{eqnarray}
\frac{d\sigma }{dy} &=&\pm 2\,b_{1}\,\sigma \,\sqrt{1+c_{0}\sigma
^{b_{1}/c_{1}}-\frac{c_{1}}{b_{1}-2c_{1}}\sigma ^{2}},\text{ }  \label{39} \\
\text{for }(b_{1} &\neq &2c_{1}),  \notag \\
&&  \notag \\
\frac{d\sigma }{dy} &=&\pm 2\,b_{1}\,\sigma \,\sqrt{1+\sigma ^{2}[\ln (\chi
)+d_{1}]},\text{ }  \label{40} \\
\text{for }(b_{1} &=&2c_{1}).  \notag
\end{eqnarray}%
As shown in Ref. \cite{ref52}, Eq. (\ref{39}) can be solved analytically, at
least, for four different particular combinations of constants. Moreover,
requiring that the solutions remain globally finite implies that $c_{0}$
cannot exceed some critical values. Therefore, borrowing the solutions from
Ref. \cite{ref52}, one can write the corresponding classical field, at order 
$\mathcal{\epsilon }$, for the following cases:

\vspace{0.4cm} \noindent \textbf{A1. For }$c_{0}<-2$\textbf{\ and }$%
b_{1}=c_{1}$,%
\begin{eqnarray}
&&\left. \phi _{A}^{(1)}(y,\tau )=\epsilon \left[ \frac{\left( \sqrt{%
c_{0}^{2}-4}\right) \sinh (2c_{1}y)}{\left( \sqrt{c_{0}^{2}-4}\right) \cosh
(2c_{1}y)-c_{0}}\right] \times \right.  \notag \\
&&\left. \cos (\sqrt{2\kappa _{1}}\tau )+\mathcal{O}(\mathcal{\epsilon }%
^{3}),\right.  \label{11.11} \\
&&\left. \chi _{A}^{(1)}(y,\tau )=\epsilon \left[ \frac{2}{\left( \sqrt{%
c_{0}^{2}-4}\right) \cosh (2c_{1}y)-c_{0}}\right] \times \right.  \notag \\
&&\left. \sin (\sqrt{2\kappa _{2}}\tau )+\mathcal{O}(\mathcal{\epsilon }%
^{3}).\right.  \label{12.11}
\end{eqnarray}

\noindent \textbf{A2. For }$b_{1}=4c_{1}$\textbf{\ and }$c_{0}<1/16$, 
\begin{eqnarray}
&&\left. \phi _{A}^{(2)}(y,\tau )=\epsilon \left[ \frac{\left( \sqrt{%
1-16c_{0}}\right) \sinh (4c_{1}y)}{\left( \sqrt{1-16c_{0}}\right) \cosh
(4c_{1}y)+1}\right] \times \right.  \notag \\
&&\left. \cos (\sqrt{2\kappa _{1}}\tau )+\mathcal{O}(\mathcal{\epsilon }%
^{3}),\right.  \label{13.22} \\
&&\left. \chi _{A}^{(2)}(y,\tau )=\epsilon \left[ -\frac{2}{\sqrt{\left( 
\sqrt{1-16c_{0}}\right) \cosh (4c_{1}y)+1}}\right] \times \right.  \notag \\
&&\left. \sin (\sqrt{2\kappa _{2}}\tau )+\mathcal{O}(\mathcal{\epsilon }%
^{3}).\right.  \label{14.22}
\end{eqnarray}

\noindent \textbf{B1. For }$b_{1}=c_{1}$\textbf{\ and }$c_{0}=-2$, 
\begin{eqnarray}
&&\left. \phi _{B}^{(1)}(y,\tau )=\epsilon \left[ -\frac{1}{2}\left[ \tanh
(c_{1}y)\mp 1\right] \right] \times \right.  \label{15.11} \\
&&\left. \cos (\sqrt{2\kappa _{1}}\tau )+\mathcal{O}(\mathcal{\epsilon }%
^{3}),\right.  \notag \\
&&\left. \chi _{B}^{(1)}(y,\tau )=\epsilon \left[ \frac{1}{2}\left[ 1\pm
\tanh (c_{1}y)\right] \right] \times \right.  \label{16.11} \\
&&\left. \sin (\sqrt{2\kappa _{2}}\tau )+\mathcal{O}(\mathcal{\epsilon }%
^{3}).\right.  \notag
\end{eqnarray}

\noindent \textbf{B2. For }$b_{1}=4c_{1}~$\textbf{and }$c_{0}=1/16$, 
\begin{eqnarray}
&&\left. \phi _{B}^{(2)}(y,\tau )=\epsilon \left[ \frac{1}{2}\left[ \pm
1-\tanh (2c_{1}y)\right] \right] \times \right.  \label{17.22} \\
&&\left. \cos (\sqrt{2\kappa _{1}}\tau )+\mathcal{O}(\mathcal{\epsilon }%
^{3}),\right.  \notag \\
&&\left. \chi _{B}^{(2)}(y,\tau )=\epsilon \left[ \sqrt{2}{\ }\frac{\cosh
(c_{1}y)\pm \sinh (c_{1}y)}{\sqrt{\cosh (2c_{1}y)}}\right] \times \right.
\label{18.22} \\
&&\left. \sin (\sqrt{2\kappa _{2}}\tau )+\mathcal{O}(\mathcal{\epsilon }%
^{3}).\right.  \notag
\end{eqnarray}

The solutions of types A1 -- B2 are illustrated in Figs. \ref%
{fig2:oscillatingkinks} and \ref{fig3:oscillons} that display the\ $\phi $
and $\chi $ components for solutions of type $A1$ and a particular set of
the parameters, as indicated in the figure caption. The solutions in the $%
\phi $ component interpolate at $|y|\rightarrow \infty $ between different
vacua of the classical-field potential. Their spatial profiles look like
kinks in the $\phi ^{4}$ theory, hence we call them oscillating kinks. The
localized $\chi $ components will be called oscillating lumps.

\begin{figure}[h]
\begin{center}
\includegraphics[width=8.2cm]{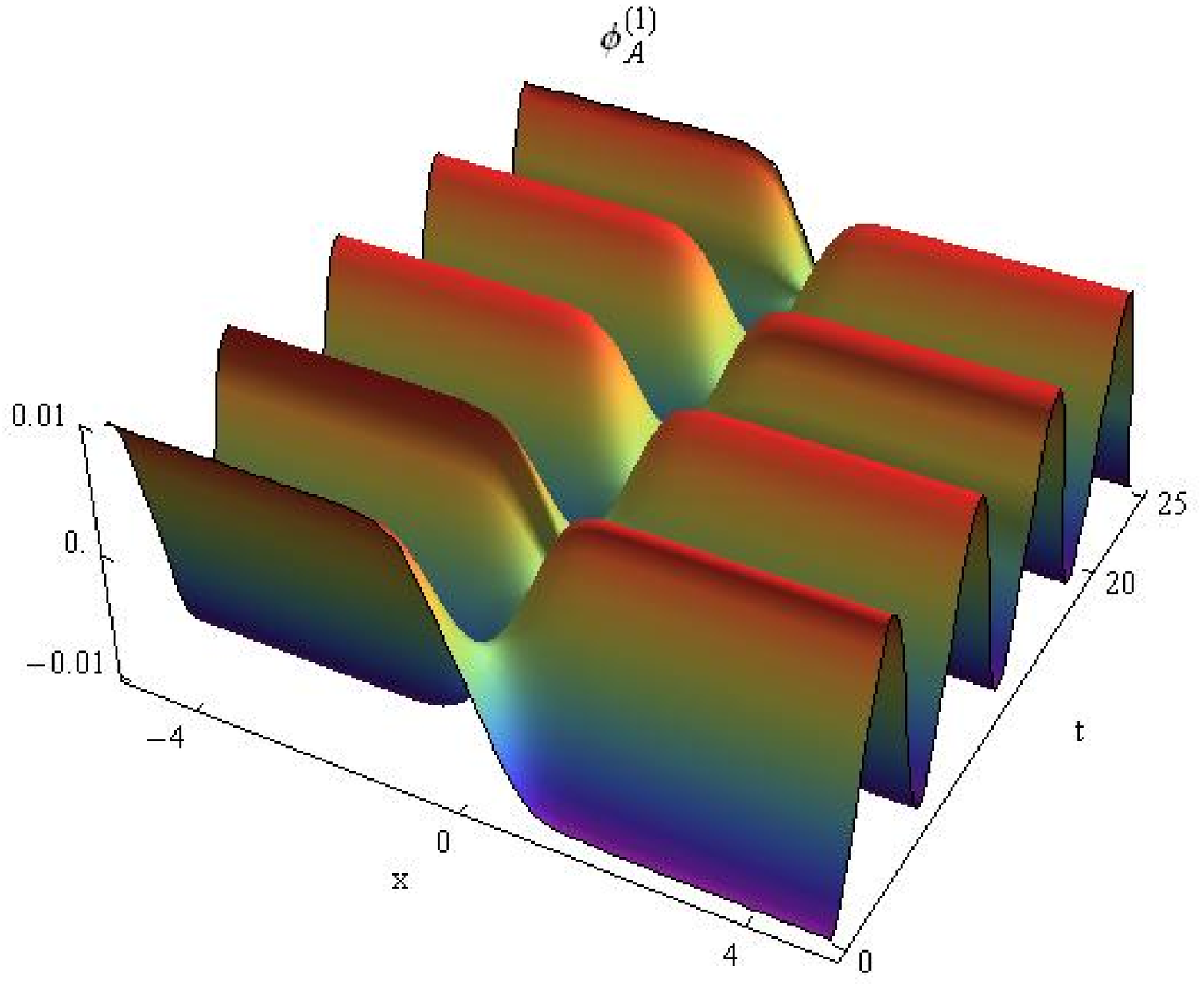} %
\includegraphics[width=8.2cm]{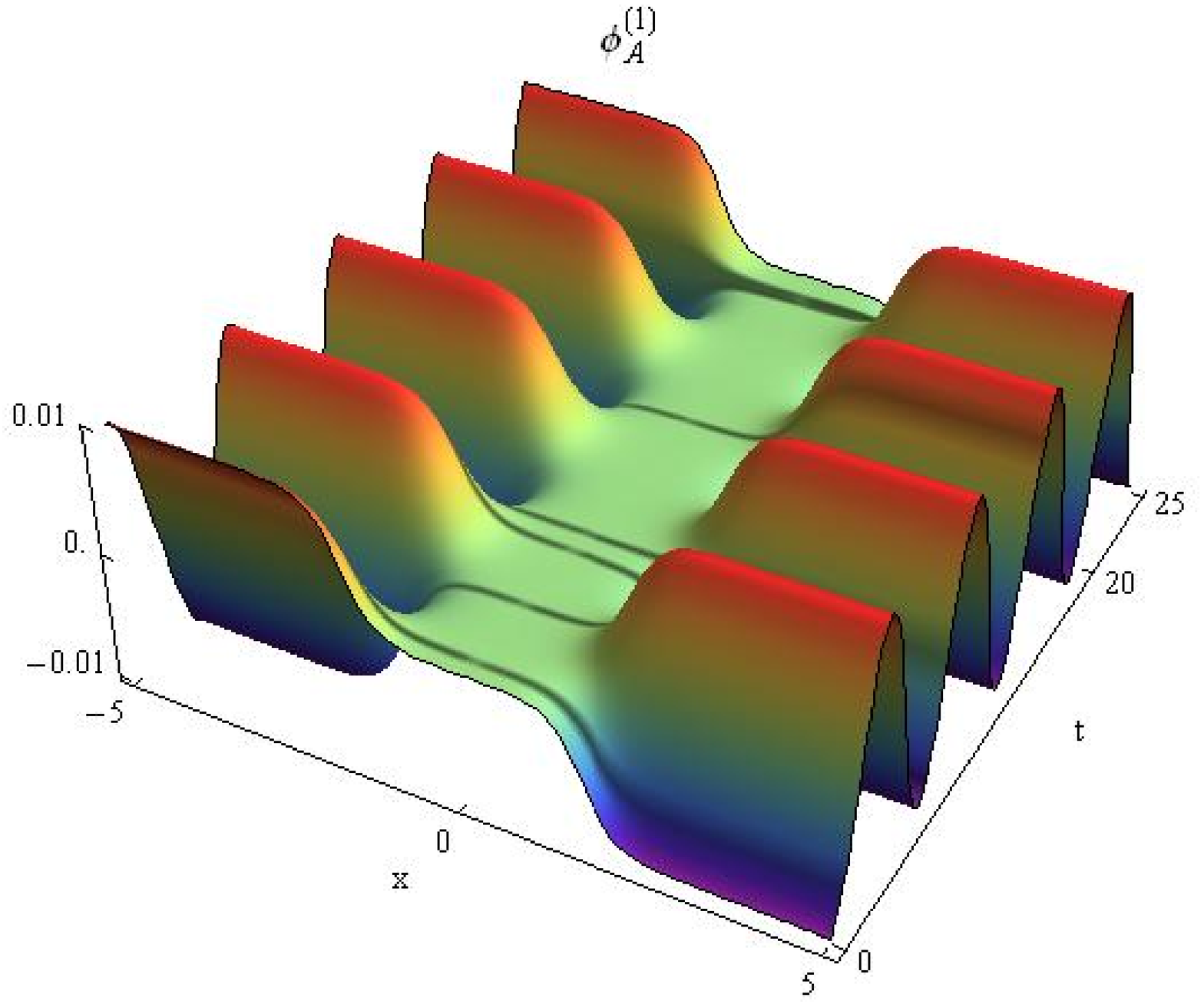}
\end{center}
\caption{Oscillating kinks computed for $\protect\lambda =\protect\mu =-1$, $%
\protect\nu =1$, $\protect\epsilon =0.01$ and $\protect\delta =0.6$. Note
that we are using the potential from Eq. (\protect\ref{1}), whose parameters
are related to those in Eq. (\protect\ref{eq0.0}) by the relations defined
in Eq. (\protect\ref{param}). The top and bottom figures show the solutions,
respectively, for $c_{0}=-2.1$ and $c_{0}=-2.00001$.}
\label{fig2:oscillatingkinks}
\end{figure}

\begin{figure}[h]
\begin{center}
\includegraphics[width=8.2cm]{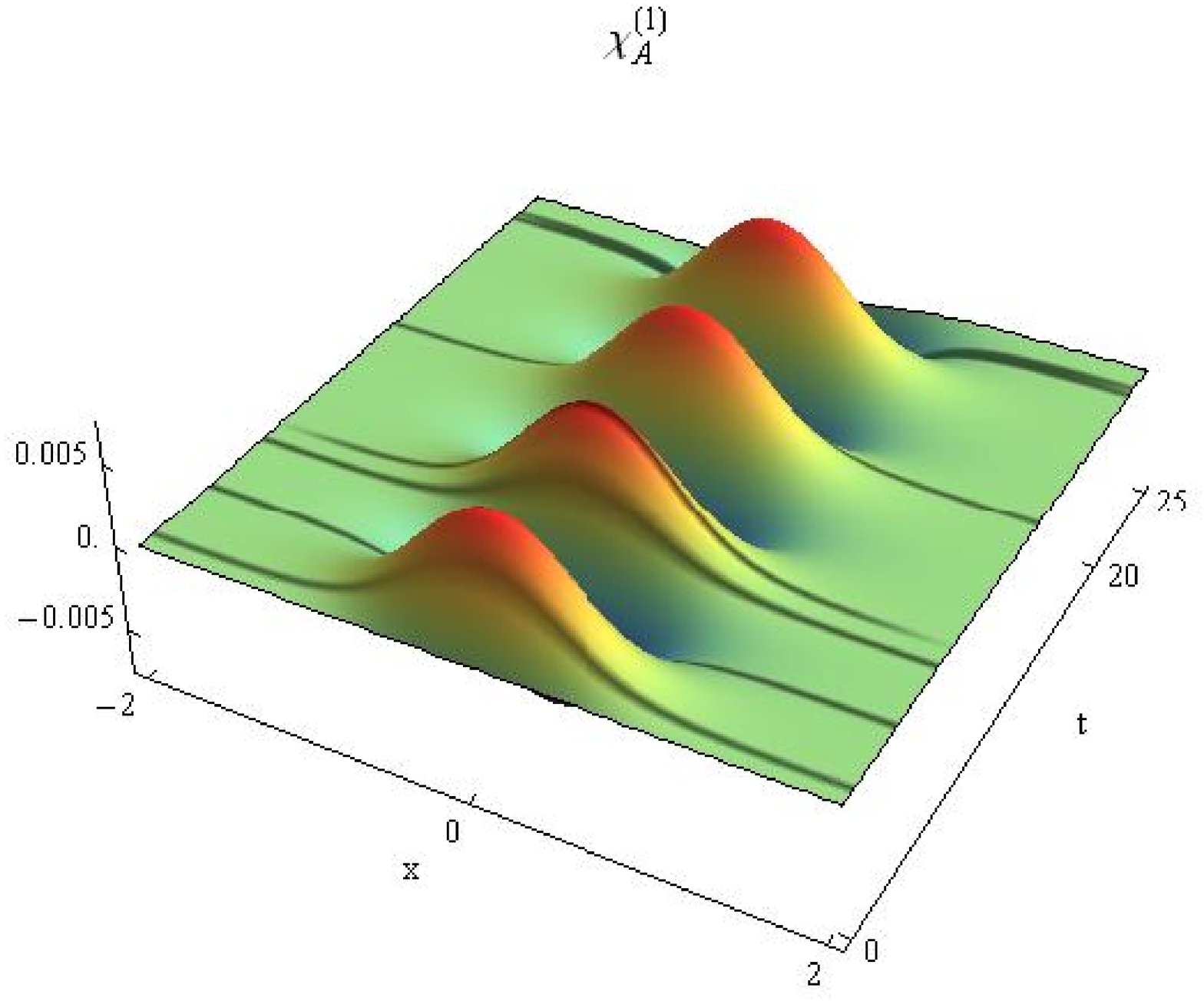} %
\includegraphics[width=8.2cm]{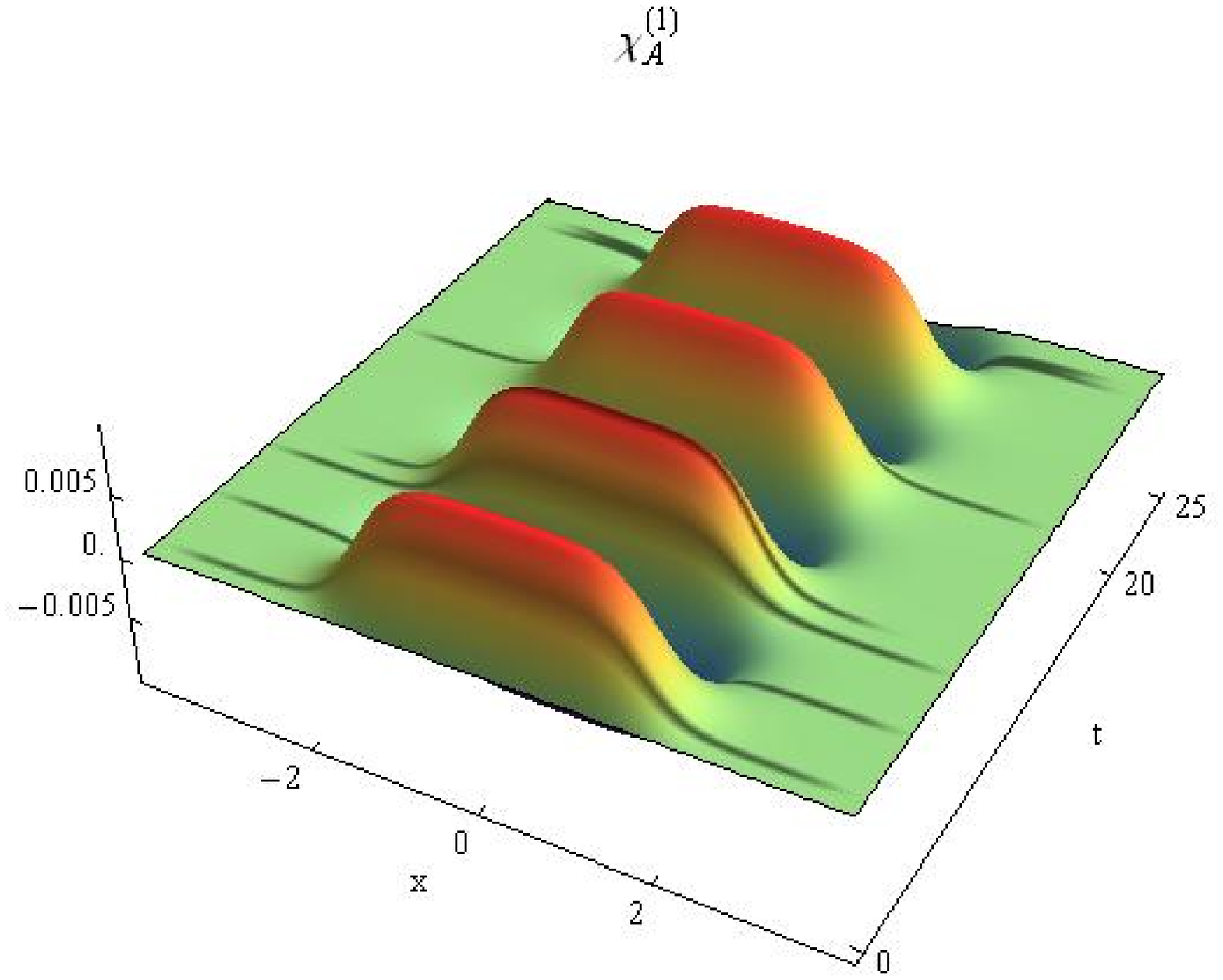}
\end{center}
\caption{Oscillon configurations for the same parameters as in the top and
bottom panels of Fig. \protect\ref{fig2:oscillatingkinks}.}
\label{fig3:oscillons}
\end{figure}

The displayed examples demonstrate that the system with two scalar fields
admits approximate solutions that oscillate with time, with the $\phi $
component keeping the oscillating kink-like profile, while the $\chi $
component features the oscillating lump-shape. Note that in all the families
of solutions, the presence of a nonvanishing arbitrary parameter $c_{1}$ is
necessary for the emergence of a flat segment in the lump (the $\chi $
field) and double-kink structure in the $\phi $ component.

\section{Emission of radiation}

As shown in the seminal work by Segur and Kruskal \cite{kruskal} (see also
Ref. \cite{KM}), one-dimensional oscillons very slowly decay through
emission of small-amplitude radiation waves. In this section we compute the
outgoing radiation of the oscillons and oscillating kinks, by means of a
method similar to the one elaborated by Hertzberg \cite{hertz} for a single
scalar field. Here we extend it for systems with two scalar fields.

The method relies on writing the solutions of the classical field equations
in the following form:

\begin{gather}
\phi _{\mathrm{osc}}(x,t)=\phi _{\mathrm{kink}}(x,t)+\xi _{\mathrm{rad}%
}(x,t),  \label{rad1} \\
\chi _{\mathrm{sol}}(x,t)=\chi _{\mathrm{osc}}(x,t)+\zeta _{\mathrm{rad}%
}(x,t),  \label{rad2}
\end{gather}%
where $\phi _{\mathrm{osc}}$ and $\chi _{\mathrm{sol}}$ are the
oscillating-kink and oscillon solutions, respectively, while, $\xi _{\mathrm{%
rad}}$ and $\zeta _{\mathrm{rad}}$ represents small radiation components,
which are generated by linearized equations following from the substitution
of the decompositions (\ref{rad1}) and (\ref{rad2}) in Eqs. (\ref{4}) and (%
\ref{5}):

\begin{eqnarray}
\square \xi _{\mathrm{rad}}+\Gamma _{\xi }\xi _{\mathrm{rad}} &=&-j_{\xi
}(x,t),  \label{rad10} \\
\square \zeta _{\mathrm{rad}}+\Gamma _{\zeta }\zeta _{\mathrm{rad}}
&=&-j_{\zeta }(x,t),  \label{rad11}
\end{eqnarray}%
where $\square$ stands the usual D'Alembertian operator, $\Gamma _{\xi
}\approx 2\kappa _{1}$ and $\Gamma _{\zeta }\approx 2\kappa _{2}$, and the
radiations sources are written as

\begin{eqnarray}
j_{\xi }(x,t) &=&\epsilon ^{3}\kappa _{3}\varphi ^{3}\cos (3\omega \sqrt{%
2\kappa _{1}}t)+...,  \label{rad12} \\
j_{\zeta }(x,t) &=&\epsilon ^{3}\kappa _{4}\sigma ^{3}\sin (3\omega \sqrt{%
2\kappa _{2}}t)+...~.  \label{rad13}
\end{eqnarray}%
%
%
%
%
%
%
%

Following the approach developed in Ref. \cite{hertz}, we can write the
radiation fields in the form

\begin{eqnarray}
\xi _{\mathrm{rad}}(x,t) &=&-\frac{1}{(2\pi )^{2}}\int \int dKd\Omega \times
\label{rad16} \\
&&\frac{j_{\xi }(K,\Omega )}{K^{2}-\Omega ^{2}+1\pm i0^{+}}e^{i(Kx-\Omega
t)},  \notag \\
\zeta _{\mathrm{rad}}(x,t) &=&-\frac{1}{(2\pi )^{2}}\int \int dKd\Omega
\times  \label{rad17} \\
&&\frac{j_{\zeta }(K,\Omega )}{K^{2}-\Omega ^{2}+1\pm i0^{+}}e^{i(Kx-\Omega
t)}.  \notag
\end{eqnarray}

In the above equation, $j_{\xi }(K,\Omega )$ and $j_{\zeta }(K,\Omega )$ are
the Fourier transforms of $j_{\xi }(x,t)$ and $j_{\zeta }(x,t)$. Then, after
straightforward manipulations, we obtain

\begin{eqnarray}
&&\left. \xi _{\mathrm{rad}}(x,t)=-\frac{\epsilon ^{3}\kappa _{3}}{8\pi ^{2}%
\sqrt{\bar{n}_{1}^{2}-\Gamma _{\xi }}}\cos (K_{\mathrm{rad}}^{(1)}x+\Delta
_{1})\times \right.  \label{rad18} \\
&&\left. \cos (\bar{n}_{1}t)j_{\xi }(K_{\mathrm{rad}}^{(1)}),\right.  \notag
\\
&&\left. \zeta _{\mathrm{rad}}(x,t)=-\frac{\epsilon ^{3}\kappa _{4}}{8\pi
^{2}\sqrt{\bar{n}_{2}^{2}-\Gamma _{\zeta }}}\cos (K_{\mathrm{rad}%
}^{(2)}x+\Delta _{2})\times \right.  \label{rad19} \\
&&\left. \sin (\bar{n}_{2}t)j_{\xi }(K_{\mathrm{rad}}^{(2)}),\right.  \notag
\end{eqnarray}%
where we define 
\begin{eqnarray}
\bar{n}_{1} &\equiv &3\omega \sqrt{2\kappa _{1}},\text{ }K_{\mathrm{rad}%
}^{(1)}\equiv \sqrt{\bar{n}_{1}^{2}-\Gamma _{\xi }}, \\
&&  \notag \\
\bar{n}_{2} &\equiv &3\omega \sqrt{2\kappa _{2}},\text{ }K_{\mathrm{rad}%
}^{(2)}\equiv \sqrt{\bar{n}_{2}^{2}-\Gamma _{\zeta }},
\end{eqnarray}%
and $\Delta _{1}$ and $\Delta _{2}$ being phase constants.






\begin{figure}[h]
\begin{center}
\includegraphics[width=8.2cm]{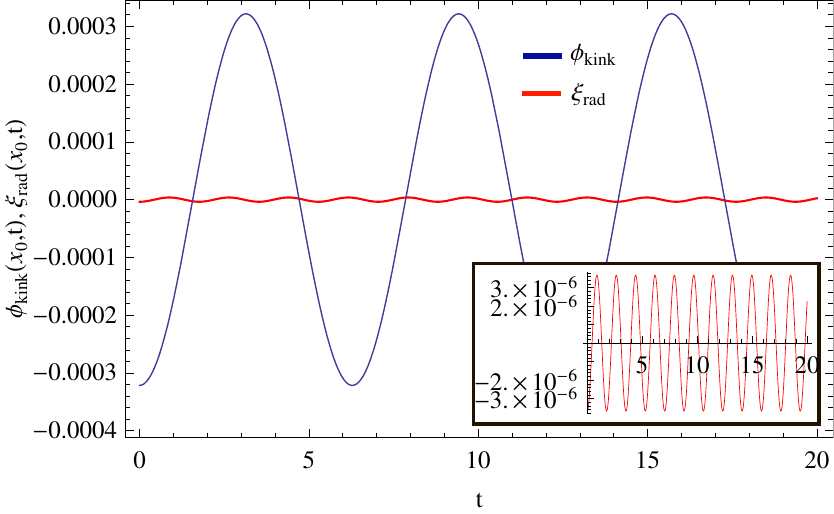} %
\includegraphics[width=8.2cm]{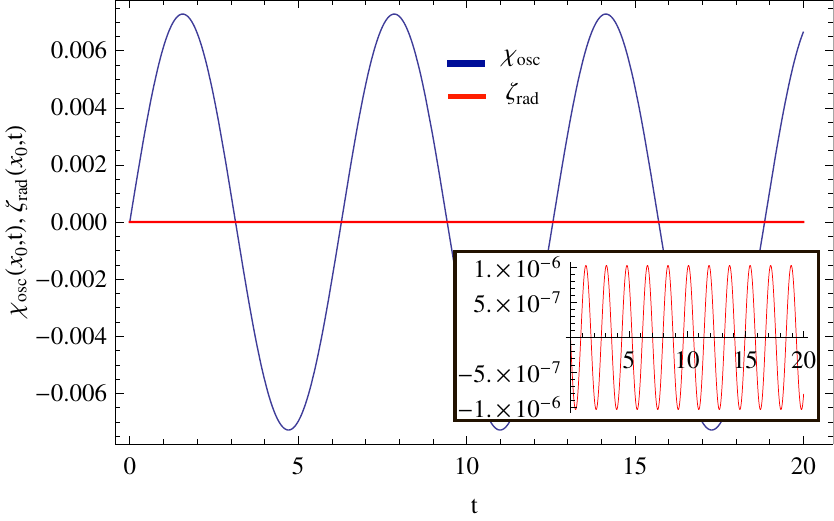}
\end{center}
\caption{The amplitude of the outgoing radiation determined by the Fourier
transform for solutions with $b_{1}=c_{1}$ and with $c_{0}<-2$. The top and
bottom plots shows the solutions computed with $c_{0}=-2.1$. The potential (%
\protect\ref{1}) is used here, with $\protect\lambda =\protect\mu =-1$, $%
\protect\nu =1$, $\protect\epsilon =0.01$ and $\protect\delta =0.6$.}
\label{fig4:radiation}
\end{figure}

In Fig. \ref{fig4:radiation} we show how the amplitudes of the outgoing
radiation vary in time, being, indeed, very small in comparison with the
amplitudes of the underlying oscillating kink and oscillating-lump modes. 
Furthermore, we also examined how the outgoing radiation vary in time for
the configurations corresponding to Eqs. (\ref{13.22}) and (\ref{14.22}),
concluding that the qualitative picture remain the same.

It has been checked by means of long simulations (see below) that the
solutions classified above as A1 an A2 are completely stable. \textcolor{black}{At this point, it is important to remark that the physical reason for their stability comes from the fact that when $c_0<-2$ and $c_0<1/16$ the spatial profiles of the solutions belong to the range of lower configurational entropy (CE) \cite{rafael-gleiser-PLB}, thereby ensuring the stability of the structures for a long period of time, but not infinitely.} On the other
hand, the solutions of the B1 type are unstable, and B2 are partially
stable. Namely, in solutions B1 both components are unstable, while in B2
the field $\phi $ collapses very fast, while component $\chi $ is a
long-standing one. Thus, it is concluded that B1 solutions decay very fast
after their creation, and in B2 the two-field configuration rapidly evolves
into a long-lived single-field oscillon, while the A1 and A2 species are
truly long-lived oscillating patterns.

\section{Numerical Results}

\begin{figure}[h]
\begin{center}
\includegraphics[width=8.2cm]{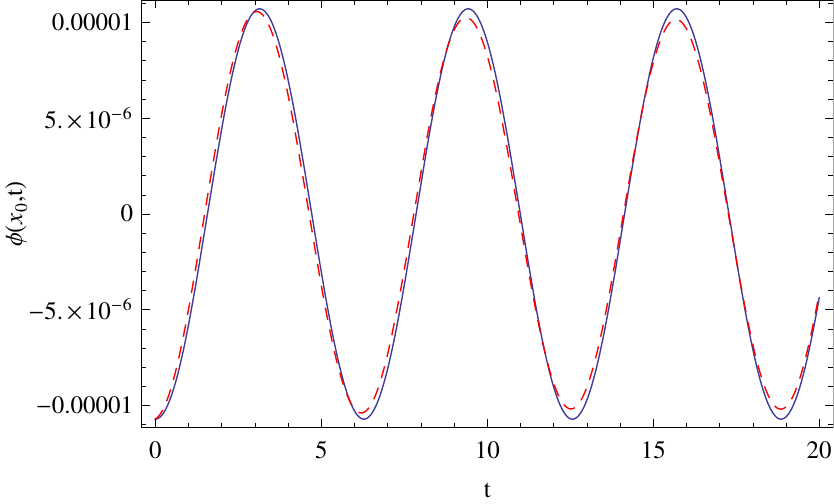} %
\includegraphics[width=8.2cm]{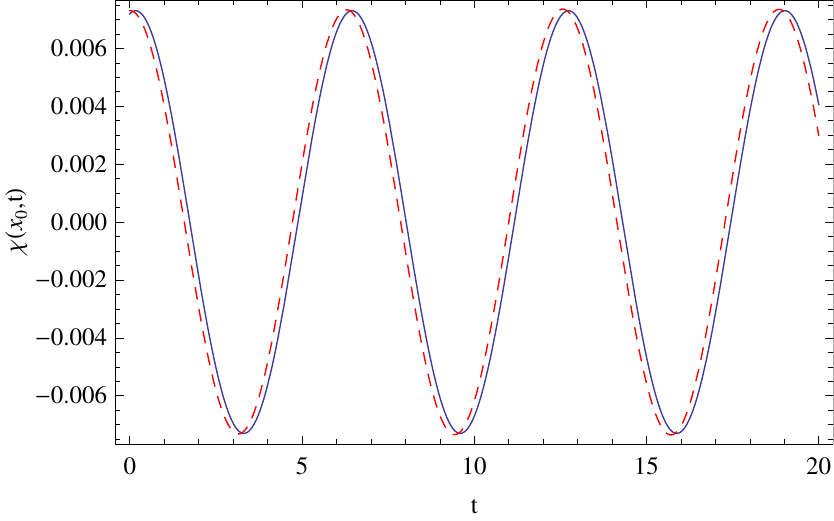}
\end{center}
\caption{Comparison of the analytical and numerical results for $b_{1}=c_{1}$
and $c_{0}=-2.1$ at point $x_{0}=0.5$, shown, respectively, by the solid and
dashed curves. Here, the potential (\protect\ref{1}) is used, with $\protect%
\lambda =\protect\mu =-1$, $\protect\nu =1$, $\protect\epsilon =0.01$ and $%
\protect\delta =0.6$.}
\label{numerical}
\end{figure}

To check the above analytical results, numerical solutions have been
produced for the oscillons and oscillating kinks. To this end, Eqs. (\ref{4}%
) and (\ref{5}) were solved with initial conditions 
\begin{eqnarray}
\phi _{\mathrm{num}}^{(1)}(x,0) &=&\epsilon \frac{\left( \sqrt{c_{0}^{2}-4}%
\right) \sinh (2c_{1}\epsilon x)}{\left( \sqrt{c_{0}^{2}-4}\right) \cosh
(2c_{1}\epsilon x)-c_{0}},  \label{nun1} \\
\chi _{\mathrm{num}}^{(1)}(x,0) &=&\epsilon \frac{2}{\left( \sqrt{c_{0}^{2}-4%
}\right) \cosh (2c_{1}\epsilon x)-c_{0}},  \label{nun2}
\end{eqnarray}%
which correspond to coefficients $b_{1}=c_{1}$, see Eqs. (\ref{eq714}) and (%
\ref{rq715}), and $c_{0}=-2.1$. To illustrate the field configurations
produced by the simulations for oscillating kinks and oscillons, in Fig. (%
\ref{numerical}) we plot fields $\phi (x,t)$ and $\chi (x,t)$ at $x=0.5$.
The figure, as well as additional numerical results, not shown here,
demonstrate good agreement of the analytical solutions with their numerical
counterparts.

\section{Summary and conclusions}

In the present work we have investigated solutions of a system of two
nonlinearly coupled scalar fields in the one-dimensional space, with a
multiple vacuum structure. The analysis shows that systems of this type has
a rich dynamics. In particular, we were able to determine new classes of
time-dependent solutions, in the form of oscillatory kinks and lump
oscillons. As far as we know, the existence and dynamics of oscillons and
kinks in models with two scalar fields were previously addressed only in
Ref. \cite{ref35}. However, in that work only solutions with states of the
same type (kinks or lumps) in both components were considered, while here we
investigate solutions combining oscillatory kinks and lumps in the two
components. In our case, we have found that some of them \textcolor{black}{are long-lived, but not infinitely stable} (types
A1 and A2), while others (B1 and B2) are unstable. 
Despite the spatial interpolation of the classical vacua of the theory, it
is not clear if these type of solutions are stable and are representative
solutions of different topological sectors for the Manifold of the classical
solutions of the field equations. Indeed, the fact that the oscillons being
approximate solutions of the field equations also raise the question if
their time evolution can lead to a classification of the classical field
equations into topological sectors, a problem that we will address in a
future work.

The new lump and kink oscillons configurations reported here may include
relatively large flat regions, where the fields are essentially different
from the vacuum values. Away from the central flat region, the solutions
varying rapidly, approaching vacua values. Plotting the respective orbits in
the $\left( \phi ,\chi \right) $ plane, see Fig. \ref{fig3:oscillons}, one
finds that the fields surround the vacuum point, featuring very small
values, which is a typical behavior demonstrated by oscillons.

\begin{figure}[h]
\begin{center}
\includegraphics[width=8.2cm]{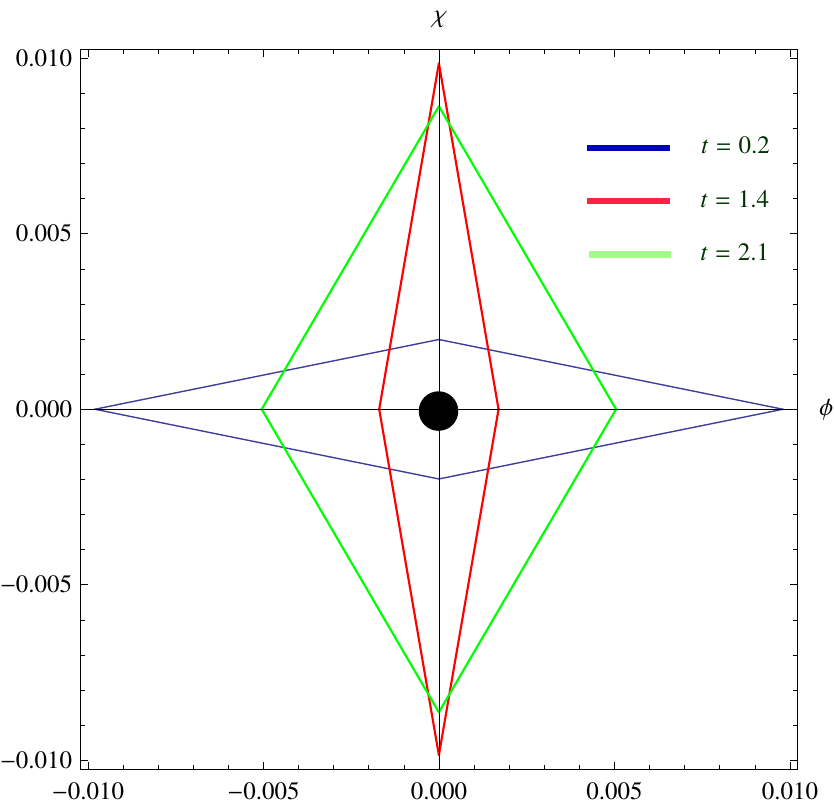} %
\includegraphics[width=8.2cm]{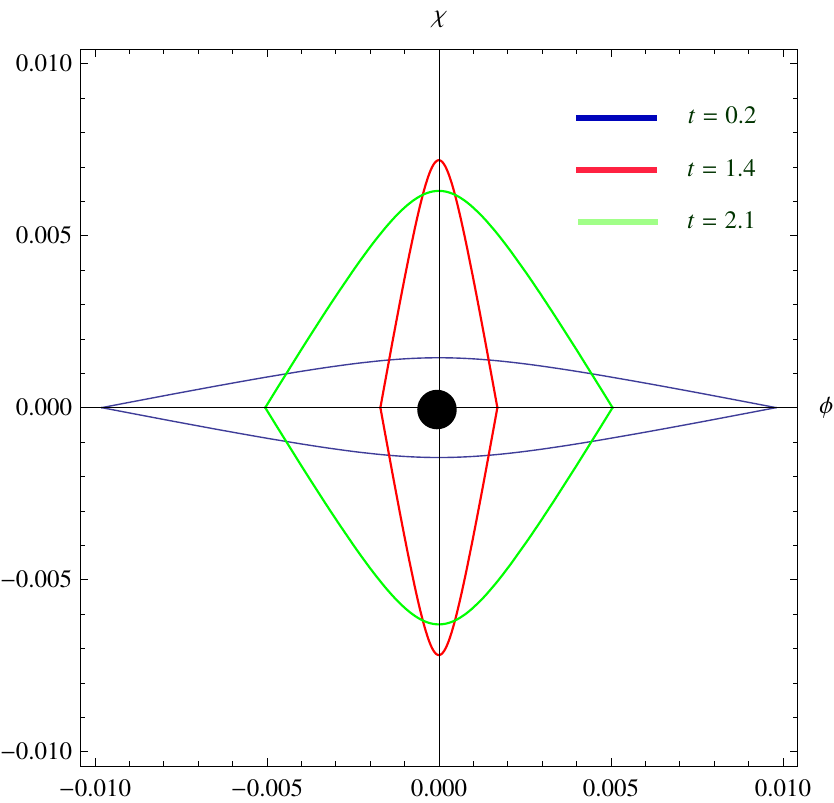}
\end{center}
\caption{Solutions in the $\left( \protect\phi ,\protect\chi \right) $ plane
for solutions with $b_{1}=c_{1}$ and vacuum state of the potential located
at point $\left( {0,0}\right) $. The top and bottom plots shows the
solutions computed with $c_{0}=-2.000001$ and $c_{0}=-2.1$, respectively. As
in Figs. \protect\ref{fig2:oscillatingkinks} and \protect\ref{fig3:oscillons}%
, these solutions pertain to $\protect\lambda =\protect\mu =-1$, $\protect%
\nu =1$, $\protect\epsilon =0.01$ and $\protect\delta =0.6$.}
\label{fig4:solutions}
\end{figure}

Our interest in the two-component system considered in this work is that it
provides a relatively simple model with a rich vacuum structure, if compared
to other theories with complex vacuum structures, such as gauge theories.
Further, the model considered in the current work has applications to
condensate matter, cosmology and high-energy physics. In particular, in the
study of low-dimensional materials similar to graphene, scalar fields
naturally appear as models of impurities and defects \cite{CastroNeto:2009zz}%
, and also model a carbon structure on the top of which the dynamics of the
electrons occurs \cite{Oliveira:2010hq}.

\label{sec:con} \vspace{1cm} \acknowledgments RACC thanks FAPESP, grants
2016/03276-5 and 2017/26646-5, for financial support. RACC would like to
thank Profs. Gyula Fodor and P\'eter Forg\'acs from Wigner Research Centre
for Physics for helpful discussion concerning oscillons. RACC also thanks
Tel Aviv University, University of Coimbra and Tokyo University of Science
for their hospitality during the development of this work. RACC also thanks
to Prof. Giuseppe Mussardo for your considerable discussions about nonlinear
field theories. The work of BAM is supported, in part, by the Israel Science
Foundation through grant No. 1287/17.

\end{document}